\documentclass[%
reprint,
superscriptaddress,
 amsmath,amssymb,
 aps
prb,
]{revtex4-2}

\usepackage{graphicx}
\usepackage{dcolumn}
\usepackage{bm}
\usepackage[colorlinks]{hyperref}
\usepackage{subfigure}
\usepackage[final]{changes}
\usepackage[section]{placeins}

\newcommand{\fixme}[1]%
   {\begingroup{\color{blue}[NOTE: \textit{#1}]}\endgroup}

\graphicspath{{Figures/},
    {Figures/two-qubit/optimized/high/},
    {Figures/two-qubit/optimized/med/},
    {Figures/two-qubit/optimized/low/},
}

\begin{document}

\title{Quadratic Models for Engineered Control of Open Quantum Systems}

\author{J.P.P. Vieira} 
\address{Interdisciplinary Centre for Mathematical Modelling and Department of Mathematical Sciences, Loughborough University, Loughborough, Leicestershire LE11 3TU, UK}
\author{A. Lazarides}
\address{Interdisciplinary Centre for Mathematical Modelling and Department of Mathematical Sciences, Loughborough University, Loughborough, Leicestershire LE11 3TU, UK}
\author{T. Ala-Nissila}
\address{MSP group, Department of Applied Physics, Aalto University, P.O. Box 15600, FI-00076 Aalto, Espoo, Finland}
\address{Interdisciplinary Centre for Mathematical Modelling and Department of Mathematical Sciences, Loughborough University, Loughborough, Leicestershire LE11 3TU, UK}

\begin{abstract}
We introduce a framework to model the evolution of a class of non-Markovian open quantum systems whose environments periodically undergo an instantaneous non-unitary evolution stage. For the special case of quadratic models, we show how this approach can generalise the formalism of repeated interactions to allow for the preservation of system-environment correlations. Furthermore, its continuous zero-period limit provides a natural description of the evolution of small systems coupled to large environments in negligibly perturbed steady states. We explore the advantages and limitations of this approach in illustrative applications to thermalisation in a simple hopping ring and to the problem of initialising a qubit chain via environmental engineering.
\end{abstract}

\maketitle

\section{Introduction}

Open quantum systems, which evolve coupled to one or more large environments, are a central concept for both the development of quantum technologies and investigation into the very foundations of quantum mechanics \cite{Breuer_Petruccione,open_quantum_book}. In practice, any working quantum device can never be completely isolated from its environment, which makes these systems ubiquitous in experiments and applications. Depending on the context, environmental coupling can either be undesirable, causing dissipation and loss of coherence and entanglement \cite{Qcomp_diss,sciencepaperonSED} or may provide a valuable resource for engineered control on quantum devices \cite{NonMarkovHeateng,QDarwin} and even extraction of quantum work \cite{Korzekwa_2016}.

Although in principle the dynamics of open quantum systems follows directly from unitary evolution of the system-environment pair according to the Schr{\"o}dinger equation, a direct approach is impractical due to the computational complexity associated with the high dimensionality of any large environment's Hilbert space. Whilst there is already a great wealth of frameworks and approximate methods for describing the evolution of open quantum systems \cite{Breuer_Petruccione,corrprojsuper,stochwavemeth,dampdriv2s,Qtransport_steadydiff,open_quantum_book,ensemble_irrev}, finding novel approaches which complement existing ones remains an active field of investigation \cite{corrpic}.
Instead of coarse-graining or integrating over the environmental degrees of freedom to reduce the complexity of the problem, an interesting indirect approach to take into account the effect of large quantum environments
is the framework of repeated interactions. In this approach, the environment
is modelled by an infinite collection of identical independent quantum
systems which take turns in interacting with the system of interest \cite{doi:10.1142/S1230161210000242,wald2024stochasticresettingdiscretetimequantum}. In the case of stochastic resetting, it has been shown that such systems can be also mapped to stochastic master equations \cite{doi:10.1142/S1230161210000242}. 
Although this prescription is only physically exact for the type of scenarios
where the system of interest interacts with particle beams or with
repeatedly-reset (measuring) devices \cite{doi:10.1142/S1230161210000242}, it
can be more generally viewed as a model for a class of open environments whose behaviour is well approximated by tractable non-unitary dynamics \cite{doi:10.1142/S1230161210000242,open_repeated,rep_int_thermod,wald2024stochasticresettingdiscretetimequantum}.

The obvious limitation of the repeated interactions approach is that
by construction it cannot apply when a
considerable amount of system-environment correlations persist for
significant periods of time. Overcoming this limitation should enable progress in our understanding of fundamental problems such as the emergence of thermalisation \cite{entangle_found,PhysRevA.43.2046,PhysRevE.50.888,helbig2024theoryeigenstatethermalisation}, as well as technological challenges such as fast qubit initialisation protocols \cite{qubit_init,tunenv,prot_reset,q_refrig_tunnel}.

In this work, we introduce a more general approach to modelling non-unitary
effects of large environments, based on periodically resetting the
state of the environment -- which reduces to repeated interactions
in one of two physically relevant limits on which we focus here. Crucially,
the other relevant limit of this technique allows for system-environment
correlations to evolve freely. We show how exact results can be analytically
obtained, for finite times between iterations and in the continuous
limit, for the special case of quadratic Hamiltonian models (in which
repeated interactions can be solved \cite{repint_continuous,RepInt,repeated_int_chaos_game,kumar2026coherencecontrolledquantumzenodynamics,rjlr-7y2k,kumar2026thermodynamicscoherenceselectivequantumreset,kumar2026memorytransportcompetitionphysicaltwobath}).

In Section \ref{sec:Quadratic-Models} we review some well-known results on quadratic models and explore in what sense they can be said to thermalise.
In Section \ref{sec:Resetting-Procedure} we introduce the resetting
procedure for quadratic models, and show how it can be analytically
solved (up to a system of linear equations). In Section \ref{sec:Examples-of-Application}
we illustrate this approach by applying it to specific situations
of interest, while comparing the repeated interactions limit to one
in which system-environment correlations evolve freely. Finally, in
Section \ref{sec:Conclusions} we take stock of the advantages and
limitations of this approach, and discuss future directions for its
applications.

\section{Quadratic Models\label{sec:Quadratic-Models}}

\subsection{Exact solutions}

In this work we focus on quadratic models, described by Hamiltonians
of the form
\begin{equation}
\hat{H}=\sum_{\alpha\beta}\hat{a}_{\alpha}^{\dagger}M_{\alpha\beta}\hat{a}_{\beta},\label{eq:quadH}
\end{equation}
where the indices $\alpha$ and $\beta$ run between $0$ and some
(large) integer $N-1$; $\hat{a}_{\alpha}^{\dagger}$ and $\hat{a}_{\beta}$
denote either fermionic or bosonic creation and annihilation operators,
respectively; and $M_{\alpha\beta}$ are the elements of a Hermitian
square matrix. This Hamiltonian can be thought of as modelling particle hopping in a lattice (the topology of which is set by the connectivity of $M_{\alpha\beta}$). Such models, by virtue of their quadratic form and of the canonical commutation relations their operators obey, are exactly and straightforwardly solvable as shown in detail in Appendix~\ref{sec:quadratic-models}. In particular, states are fully specified by their corresponding \emph{single-particle density matrix}, 
$\rho_{\alpha\beta}\equiv\left\langle \hat{a}_{\alpha}^{\dagger}\hat{a}_{\beta}\right\rangle$, whose time evolution is fully determined by $M_{\alpha\beta}$. Similarly, many-body expectation values involving more than two operators $a_\alpha,a^\dagger_\beta$ are also uniquely defined from the single-particle density matrix via Wick's theorem (see, e.g., Ref. \onlinecite{Lazarides:2014cl}).

\subsection{Pseudo-thermalisation in open quadratic systems\label{subsec:Equilibration-and-pseudo-thermal}}

Quadratic models present a unique opportunity for the study of open quantum systems. Interpreting $\hat{a}_{\alpha}^{\dagger}$ as creating
a particle with a well-defined position at the $\left(\alpha+1\right)^{{\rm th}}$
site in a lattice, we can refer to a small (typically contiguous) part of this lattice as belonging to a system ($\alpha\in\mathcal{S}$) and to the rest of it as being in an environment ($\alpha\in\mathcal{E}$). Therefore we can treat this small system as a model of an open system as long as we can solve the quadratic problem for a sufficiently large environment. The simple dynamics of quadratic models combined with the fact that the state can be encoded in an $N\times N$ matrix (the single-particle density matrix--this is to be contrasted to requiring a $2^{N}\times 2^{N}$ density matrix in the fermionic case, and a potentially infinite one in the bosonic case), this approach permits us to computationally reach larger environment sizes than using more complex models.

Integrable models (of which quadratic models such as ours are a simple example) equilibrate after long times under unitary evolution in the thermodynamic limit~\cite{staquench}, i.e., time-averaged quantities tend to a constant in the infinite future. In fact, for all quadratic models without specific relations  between energy levels~\cite{Khemani2016,Sreejith:2016wz,Lazarides2017a}, the time-averaged single-particle density matrix in this limit reduces to the diagonal ensemble result
\begin{equation}
\begin{split}
\overline{\rho}_{\alpha\beta}
    &\equiv\lim_{T\rightarrow\infty}\frac{1}{T}\intop_{0}^{T}\rho_{\alpha\beta}\left(t\right)dt\\
    &=\sum_{n}\psi_{n\alpha}^{\ast}\psi_{n\beta}\widetilde{\rho}_{nn}\left(0\right)\\
    &=\sum_{n\alpha^{\prime}\beta^{\prime}}\psi_{n\alpha}^{\ast}\psi_{n\beta}\psi_{n\alpha^{\prime}}\psi_{n\beta^{\prime}}^{\ast}\rho_{\alpha^{\prime}\beta^{\prime}}\left(0\right),
\end{split}
\label{eq:longtime_rho}
\end{equation}
where the $\psi_{n\alpha}$ are components of the $n$th eigenvector (see Eq.~(\ref{eq:diagH}) and the discussion surrounding it). This corresponds to dropping all off-diagonal entries in the initial single-particle density matrix in the (single-particle) energy eigenbasis, $\widetilde{\rho}_{nm}$ (see Eq.~(\ref{eq:til-spdm})). For certain non-integrable models, the Eigenstate Thermalisation Hypothesis (ETH) \cite{PhysRevA.43.2046,PhysRevE.50.888} provides a mechanism by which the expectation values of local operators in large unitary systems approach thermal expectation values through interactions between their subsystems. It has been recently argued  \cite{helbig2024theoryeigenstatethermalisation} that the eigenvalue
distribution of interacting quantum systems being a Gaussian under very general circumstances directly results in the ETH holding.

Generally, we are interested in the long-time expectation values of operators supported within the system (which can be expressed as combinations of terms involving $\overline{\rho}_{\alpha\beta}$ for $\alpha,\beta \in \mathcal{S}$). In the more realistic case of non-integrable models, the generic expectation \cite{entangle_found} is that such operators not
just equilibrate in the sense of Eq.~(\ref{eq:longtime_rho}) ~\cite{Reimann:2008hq} but also thermalise in such a way that, in the thermodynamic limit (i.e., when $\#\mathcal{E}$ is arbitrarily large),
$\overline{\rho}_{\alpha\beta}$ will generically be arbitrarily close to a thermal (Gibbs) state. For integrable (and in particular free, such as here) systems, this generally is not the case~\cite{Rigol:2007bm} unless a special initial state is chosen. One therefore does not expect thermalisation in quadratic models. Nevertheless, a weaker phenomenon, which we call pseudo-thermalisation, might still occur. This is when, in the thermodynamic limit, the long-time state of the system (i.e., the section of $\overline{\rho}_{\alpha\beta}$ corresponding only to system correlations) does not depend on the initial state of the system. This is the case that we will consider in the present work.
From Eq. (\ref{eq:longtime_rho}), it is straightforward to verify that
pseudo-thermalisation takes place whenever
\begin{equation}
\lim_{N\rightarrow\infty}\sum_{n}\psi_{n\alpha}^{\ast}\psi_{n\beta}\psi_{n\gamma}\psi_{n\delta}^{\ast}=0\ \forall\alpha,\beta,\gamma,\delta\in\mathcal{S},\label{eq:pseudothermal}
\end{equation}
where the the limit for large $N$ is imposed by the thermodynamic limit \footnote{Mathematically, one can always interpret the thermodynamic limit as
a continuous limit of such lattice models, for example by defining
a sequence of Hamiltonians $\hat{H}\left[N\right]$ such that the
size of the environment relative to the size of the system grows with
increasing $N$ -- and then the limit in Eq. (\ref{eq:pseudothermal})
can be taken literally.}.

Since the eigenvectors $\psi_{n\alpha}$ are orthonormal
by construction, most models will obey Eq. (\ref{eq:pseudothermal}) \footnote{i.e., most sets of orthonormal $\psi_{n\alpha}$ can be seen to verify this property -- e.g., by checking the entries in normal matrices randomly generated by taking the imaginary exponential of uniformly randomly
picked Hermitian matrices, or by QR-decomposing such Hermitian
matrices.}. Intuitively, this follows from the
expectation that a typical element of such orthonormalised eigenvectors
will have a magnitude of order $\mathcal{O}\left({1}/{\sqrt{N}}\right)$, and thus the sum in Eq. (\ref{eq:pseudothermal}) will be of order $\mathcal{O}\left({1}/{N}\right)\longrightarrow 0$. In particular,
any translation-invariant quadratic model must pseudo-thermalise in
this way \footnote{As $\Sigma_{n}\left|\psi_{n\alpha}\right|^{2}=1$ implies $\left|\psi_{n\alpha}\right|={1}/{\sqrt{N}}$
if all sites are equivalent. In the special case of single-site systems,
the resulting relation $\Sigma_{n}\left|\psi_{n\alpha}\right|^{4}={1}/{N}$
can additionally be shown to correspond to the fastest possible convergence
to pseudo-thermalisation with increasing $N$.} (unless its energy levels are fine-tuned to prevent this \footnote{A notable such counterexample being the case of Bethe lattices, whose energy levels are highly degenerate.}).
Nevertheless, there are several counterexamples, such as models with broken translational invariance
leading to localised states (for example in Anderson localization) and thus a finite term in Eq. (\ref{eq:pseudothermal}).

\section{Resetting Process\label{sec:Resetting-Procedure}}

Ideally we would like to find settings in which subsystems of quadratic systems pseudo-thermalise more generically. Bearing in mind that most real-world systems are not integrable (and certainly not quadratic), we focus on situations where our quadratic environment is connected to some more complicated super-environment. In order to circumvent the mathematical difficulties that would come with actually calculating the evolution of such a complex system, the effect of this super-environment on our quadratic universe can be captured by some class of non-unitary evolution. Naturally, the exact form of this non-unitary evolution depends on the exact nature of the super-environment and its coupling with the quadratic sector.

A well-motivated example of such non-unitary evolution is found in the framework of repeated interactions \cite{RepInt_original,doi:10.1142/S1230161210000242,open_repeated,rep_int_thermod,wald2024stochasticresettingdiscretetimequantum}, where, in between periods of purely unitary evolution, the environment is periodically reset to its initial state. This comes about when the system interacts with periodically externally (and non-unitarily) reset devices (when the super-environment is made up of whatever mechanism is being used to reset the device) or when it  periodically comes into contact with identical copies of the same environment (when the super-environment is made up of all the copies of the environment that are not currently interacting with the system \footnote{Technically, in this case, the definition of the environment depends on exactly which copy is currently interacting with the system.}).

In this section we expand this framework to allow for more general resetting processes; which may, for example, preserve correlations between system and environment. In what follows, we consider a quadratic model as described in Section \ref{sec:Quadratic-Models},
which we can generally interpret as relating to particles in a lattice
which can be divided into a (small) system $\mathcal{S}$ and an environment $\mathcal{E}$ (as
discussed in Subsection \ref{subsec:Equilibration-and-pseudo-thermal}).
This model is allowed to evolve unitarily for periods of duration
$\tau$ separated by instants in which some elements of the full (joint system-environment) single-particle
density matrix are reset to their initial values. 

Mathematically, the only restriction on which elements can be reset and on their values after resetting is that this process should not violate the Hermiticity of $\rho$. Physically, these choices are determined by what physical process one seeks to model. Here we focus on two physically distinct  examples: repeated interactions and the case of environments which periodically and (approximately) instantaneously thermalise with some arbitrarily large super-environment. In the former, only the system-system block of $\rho$ is not reset, while the system-environment block are reset to zero. Conversely, in the latter only the environment-environment block is reset, to some appropriate thermal distribution.

\subsection{Evolution and long-time limit}
\label{sec:uniqueness-discrete}

The unitary part of the evolution of the single-particle density matrix is simply given by Eq. (\ref{eq:spdm_evol_op}). Defining $\mathcal{R}$ as the set of pairs of indices whose corresponding entries in $\rho$ get periodically reset, the evolved single-particle density matrix after one time interval $\tau$ is thus
\begin{equation}
\rho_{\alpha\beta}\left(t+\tau\right)=\sum_{\alpha^{\prime}\beta^{\prime}}U_{\alpha\alpha^{\prime}}^{\ast}\left(\tau\right)U_{\beta\beta^{\prime}}\left(\tau\right)\rho_{\alpha^{\prime}\beta^{\prime}}\left(t\right)
\label{eq:tau_increment_dynamics}
\end{equation}
for $\left(\alpha,\beta\right)\notin\mathcal{R}$ ($U$ being a matrix representation of the evolution operator -- see Eq.~(\ref{eq:evol_op})) and
\begin{equation}
\rho_{\alpha\beta}\left(t+\tau\right)=\rho_{\alpha\beta}\left(t\right)
\label{eq:tau_increment_reset}
\end{equation}
for $\left(\alpha,\beta\right)\in\mathcal{R}$.

Given the periodic nature of this setting, we focus on the stroboscopic evolution of the model at discrete times of the form $t=n\tau$ (for integer $n$). Then we can simply ignore the evolution of those elements $\rho_{\alpha\beta}$ for which $\left(\alpha,\beta\right)\in\mathcal{R}$ and organise the remaining entries in a state vector
\begin{equation}
V_{i}\left[n\right]\equiv\rho_{\alpha_{i}\beta_{i}}\left(n\tau\right),\label{eq:state_vector}
\end{equation}
where $\left(\alpha_{i},\beta_{i}\right)\notin\mathcal{R}$ is a sequence
that covers each pair of indices not in $\mathcal{R}$ exactly once. Eq. (\ref{eq:tau_increment_dynamics}) can then be rewritten
as the discrete evolution equation
\begin{equation}
V_{i}\left[n+1\right]=\sum_{j}D_{ij}V_{j}\left[n\right]+C_{i},\label{eq:lineq}
\end{equation}
where we have introduced the square (non-Hermitian) matrix
\begin{equation}
D_{ij}=U_{\alpha_{i}\alpha_{j}}^{\ast}\left(\tau\right)U_{\beta_{i}\beta_{j}}\left(\tau\right)\label{eq:D}
\end{equation}
and the constant vector
\begin{equation}
C_{i}=\sum_{\left(\alpha^{\prime},\beta^{\prime}\right)\in\mathcal{R}}U_{\alpha_{i}\alpha^{\prime}}^{\ast}\left(\tau\right)U_{\beta_{i}\beta^{\prime}}\left(\tau\right)\rho_{\alpha^{\prime}\beta^{\prime}}\left(0\right).\label{eq:C}
\end{equation}

The matrix $D$ only depends on the model Hamiltonian (through Eqs. (\ref{eq:D}) and (\ref{eq:evol_op})), whereas the constant vector $C$ additionally depends on the manner in which resetting is implemented (through $\rho_{\alpha^{\prime}\beta^{\prime}}\left(0\right)$ in Eq. (\ref{eq:C})). We therefore intuitively expect that the rate of convergence to any attractor will be mostly set by properties of the Hamiltonian (which makes no reference to the super-environment), whilst the state to which $\mathcal{R}$ is reset will play a role in shifting these attractors away from the null vector \footnote{Note that the case where the null vector is an attractor solution to Eq. (\ref{eq:lineq}) is an interesting one from the point of view of environment engineering, as discussed in some detail in Subsection \ref{subsec:initialisation}.}.

In Appendix~\ref{sec:uniqueness_of_fixed_point_of_eq_eq:lineq} we show that there is usually a unique and attractive fixed point of Eq.~(\ref{eq:lineq}), so that at long times the system pseudo-thermalises, completely forgetting its initial state.

\subsection{High-frequency resetting and the continuous resetting limit}
\label{sec:uniqueness-continuous}

One interesting limit for this procedure is when $\tau$
is much smaller than all other relevant characteristic time scales
-- particularly as $\tau=0$ corresponds to the problem of the continuous evolution
of a small system coupled to an environment in a static state (which
in particular can be taken to be a Gibbs state). 

In this continuous limit, the repeated interactions framework becomes equivalent to the Born approximation which is typically the starting point for the weak coupling limit \cite{Breuer_Petruccione}. This approximation states that the density matrix for the universe can be approximately expressed as a direct product between the system and the environment reduced density matrices, with the latter being constant. Therefore, the case we consider in which the environment is kept in a constant (thermal) state while allowing for evolving system-environment correlations can be seen as describing a less stringent version of the weak coupling limit.

In this limit,
we can expand the evolution operator to leading order in $\tau$, yielding
\begin{equation}
\begin{split}
U_{\alpha\beta}\left(\tau\right)=
    &\left\langle \alpha\right|\left(\mathbb{I}-\frac{i\hat{H}\tau}{\hbar}+\mathcal{O}\left(\tau^{2}\right)\right)\left|\beta\right\rangle\\
    &=\delta_{\alpha\beta}-\frac{i\tau}{\hbar}M_{\alpha\beta}+\mathcal{O}\left(\tau^{2}\right),
\end{split}
\label{eq:U_tauexpand}
\end{equation}

Analogously to the discrete case (see Appendix~\ref{sec:uniqueness_of_fixed_point_in_the_continuous_limit}), the continuous time evolution can then be written in the form
\begin{equation}
\frac{d\mathcal{V}_{i}}{dt}\left(t\right)=\sum_{j}\mathcal{D}_{ij}\mathcal{V}_{j}\left(t\right)+\mathcal{C}_{i},
\end{equation}
where the quantities $\mathcal{V}, \mathcal{D}$ and $\mathcal{C}$ are continuous versions of $V, D$ and $C$ in Eq.~(\ref{eq:lineq}). 
We can then show (as in Appendix~\ref{sec:uniqueness_of_fixed_point_in_the_continuous_limit}) that in this continuous limit the system also typically possesses a single (attractive) steady state, thus pseudo-thermalising.

\section{Examples of Application\label{sec:Examples-of-Application}}

\subsection{Thermalisation}

We shall now focus on situations in which the environment is reset to (or kept at) a thermal state, which offer us relatively simple physical interpretations. Where system-environment correlations are reset to zero (i.e., in the repeated interactions limit) this corresponds to the system interacting with consecutive identical independent environments at the same temperature. Where such correlations are allowed to evolve, this corresponds to a situation where the environment is actually part of a bigger ``super-environment'' with which it periodically interacts and thermalises much faster than any other relevant time scales. In the continuous limit, the latter can be seen as an approximate model of a system interacting with a thermal environment which is not noticeably perturbed by this interaction corresponding to the Born-Markov approximation on the density matrix of the environment \cite{PhysRevResearch.1.013004}.

Models for which these interactions lead to thermal attractors are of particular interest as they can be taken as toy models for thermalisation processes. Although that is not a generic property of this type of resetting procedure, we are able to find and study a few simple cases in which this takes place.

The simplest example is when the environment is repeatedly reset to an infinite temperature state, which generically induces the system to settle at a thermal attractor state with infinite temperature in both scenarios we consider (as shown in Appendix \ref{sec:inf_temp}).

This result suggests that if one is interested in finding scenarios in which quadratic models thermalise with an environment at finite temperatures, then a high-temperature limit is a natural regime to look into. As thermalisation is not a generic feature of these models away from $\beta=0$, we consider the special case of one of the simplest and most well-known quadratic models: a fermionic
periodic hopping ring, whose Hamiltonian is given by
\begin{equation}
\hat{H}=-J\sum_{\ell=0}^{N-1}\left(\hat{a}_{\ell}^{\dagger}\hat{a}_{\ell+1}+\hat{a}_{\ell+1}^{\dagger}\hat{a}_{\ell}\right),\label{eq:Hring}
\end{equation}
where $J$ is the hopping amplitude and $\hat{a}_{\ell+N}\equiv\hat{a}_{\ell}$.

To illustrate our results we focus on a ring with $N=100$ where the system is defined as a segment of eight consecutive sites, in a scenario with $\tau=0.01\hbar / J$ so as to be comfortably in the high-frequency limit \footnote{The smallest typical time scale in this ring being of order $\pi\hbar/J$. Note we deliberately avoid working in the continuous limit in this particular case due to additional mathematical complications arising from a null determinant in the result of Eq.~(\ref{eq:D_cont}).}. The evolution operator corresponding to Eq.~(\ref{eq:Hring}) (which can be straightforwardly computed using Eq.~(\ref{eq:evol_op})) can be used in Eq.~(\ref{eq:D}) and Eq.~(\ref{eq:C}) to determine the components of $D$ and $C$, respectively; which in turn can be used in Eq.~(\ref{eq:fixed_point}) (following an appropriate change of basis) to yield the elements of the system's steady-state single-particle density matrix, $\overline{\rho}^{\mathcal{S}}_{\alpha\beta}$. In Fig. 1 we show a schematic of the effect of the two types of resetting on the system-environment density matrix.

\begin{figure}[h]
\includegraphics[scale=0.195]{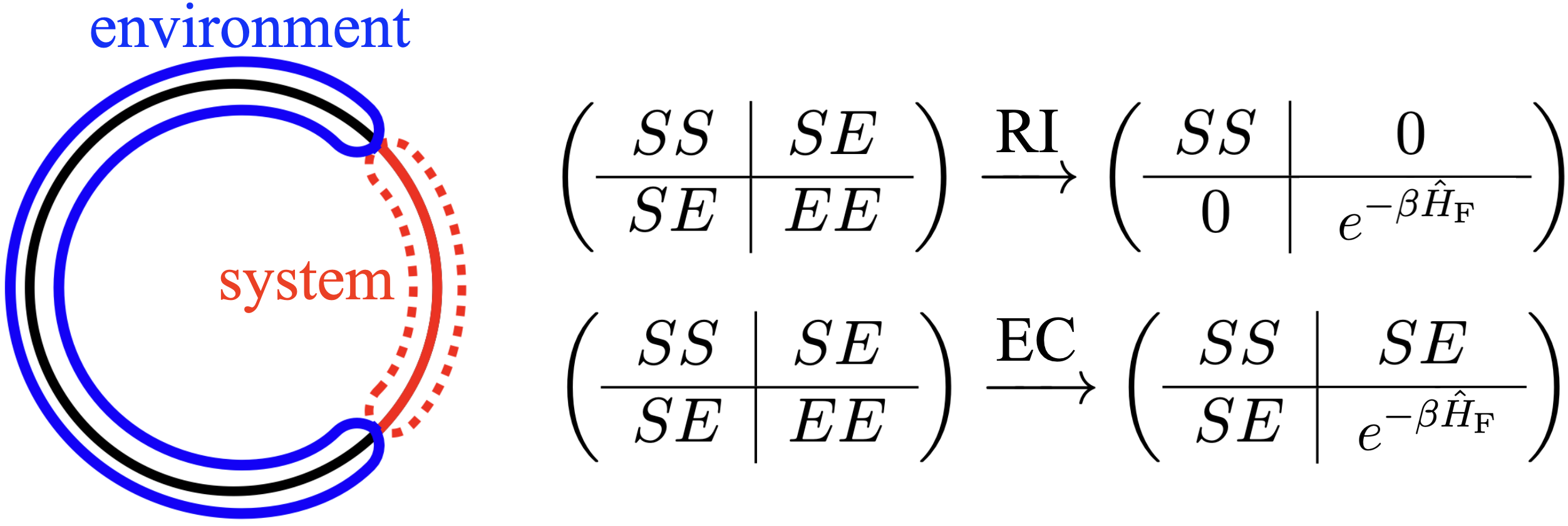}
\caption{\label{fig:fig1} A schematic of the effect of the two different forms of resetting from the super-environment. The system $S$ denoted in red is part of the total hopping ring as given by Eq. (12) and the rest in blue is the environment $E$. In the case of repeated interactions (RI), the $S-E$ matrix elements are set to zero and the $E-E$ elements adjusted to a thermal Fermi-Dirac distribution (denoted symbolically by $e^{-\beta {\hat H}_{\rm F}}$). In the case where correlations are allowed to evolve (EC), the $S-E$ elements are kept and evolve continuously.
}
\end{figure}

If the steady state of the system is approximately thermal, we should expect the corresponding single-particle density matrix to be consistent with a Fermi-Dirac distribution: being approximately diagonal in the system Hamiltonian's eigenbasis, with diagonal entries in said basis approximately corresponding to
\begin{equation}
\overline{\tilde{\rho}}^{\mathcal{S}}_{\alpha\alpha}=\frac{1}{e^{\beta E^{\mathcal{S}}_{\alpha}}+1},\label{eq:fermi_dirac}
\end{equation}
where $\beta = 1/k_{\rm B} T$ is the inverse thermal energy ($k_{\rm B}$ standing for the Boltzmann constant and $T$ for the temperature) and $E^{\mathcal{S}}_{\alpha}$ are eigenvalues of the system Hamiltonian. 

\begin{figure}[h]
\includegraphics[scale=0.14]{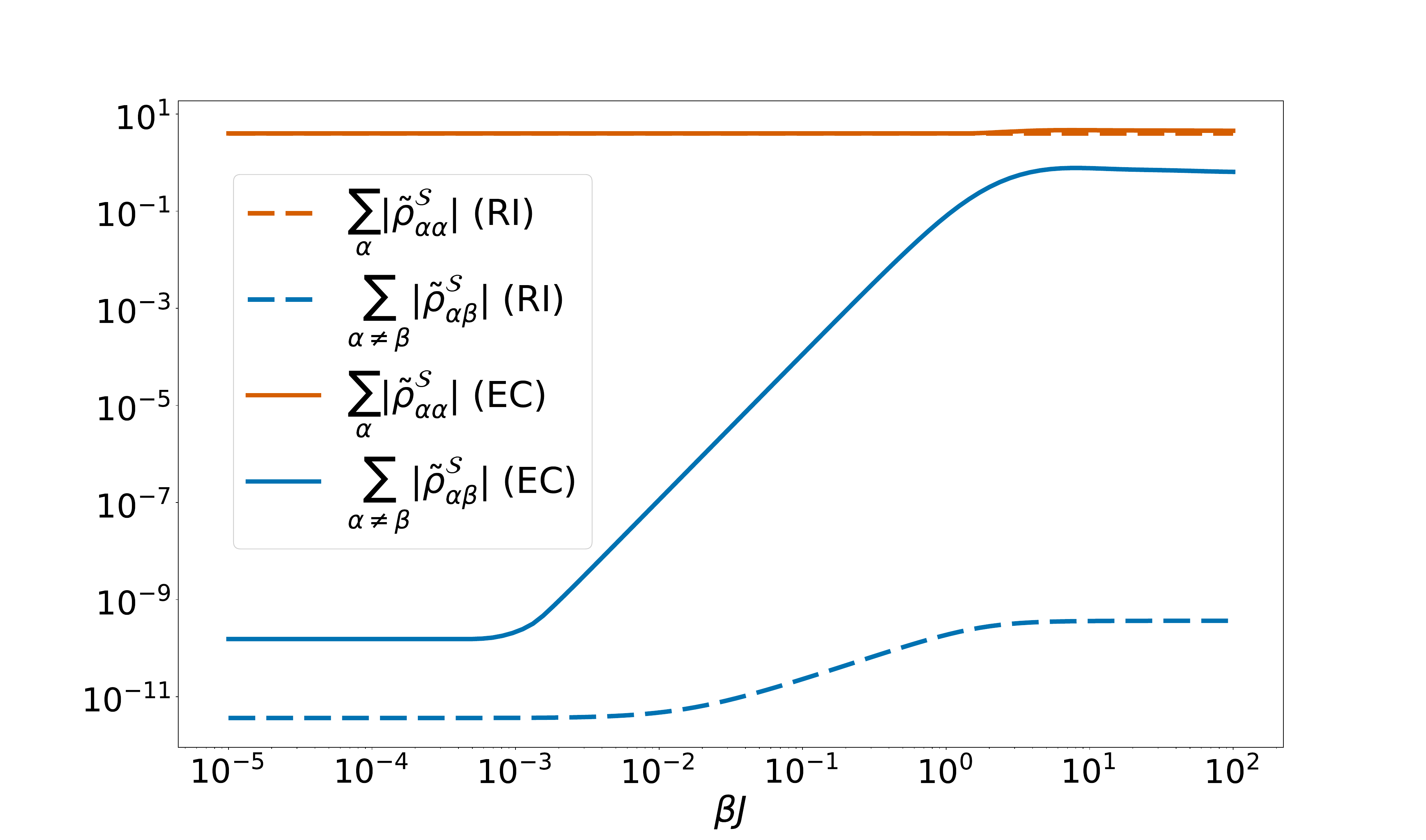}
\caption{\label{fig:diagonal_sums} Sum of modulus of diagonal (red) and off-diagonal (blue) elements of the steady-state single-particle density matrix in the system Hamiltonian's eigenbasis -- as calculated from Eq.~(\ref{eq:fixed_point}) for a system of eight sites in a 100-sites-long hopping ring coupled to an environment reset to a thermal state at temperature $\beta$ with period $\tau=0.01$, both in the repeated interactions regime (RI, dashed) and allowing system-environment correlations to evolve (EC, solid). Since an exactly thermal state has vanishing off-diagonal terms, the smallness of these terms is consistent with an approximately thermal attractor for all high temperatures and even some intermediate temperatures (or even low temperatures in the case of repeated interactions).}
\end{figure}

Figure~\ref{fig:diagonal_sums} contrasts the sums of (the modulus of) the diagonal and off-diagonal entries in $\overline{\tilde{\rho}}^{\mathcal{S}}_{\alpha\beta}$ for the case of repeated interactions and evolving system-environment correlations, as functions of the temperature of the thermal state to which the environment is reset, $\beta$. Even for relatively high values of $\beta$ (low temperatures), we can see that the weight of off-diagonal elements is negligible (especially for repeated interactions), consistent with a thermal steady state. As expected from the exact result for $\beta=0$, the higher the environment temperature the more negligible off-diagonal elements are (although there appears to be a residual constant contribution of these elements for low $\beta$, which should come from numerical errors dealing with very small numbers).

To verify the thermal nature of these steady states, it is necessary to further show that the diagonal elements follow Eq.~(\ref{eq:fermi_dirac}). This can be done by computing an effective inverse Boltzmann temperature $\beta_{\alpha}$ corresponding to each diagonal element $\overline{\tilde{\rho}}^{\mathcal{S}}_{\alpha\alpha}$ ,
\begin{equation}
\beta_{\alpha}=\frac{1}{E^{\mathcal{S}}_{\alpha}}\ln\left(\frac{1}{\overline{\tilde{\rho}}^{\mathcal{S}}_{\alpha\alpha}}-1\right)
,\label{eq:effective_fermi_dirac}
\end{equation}
and then checking that all $\beta_{\alpha}$ are approximately the same.

\begin{figure}[h]
\includegraphics[scale=0.14]{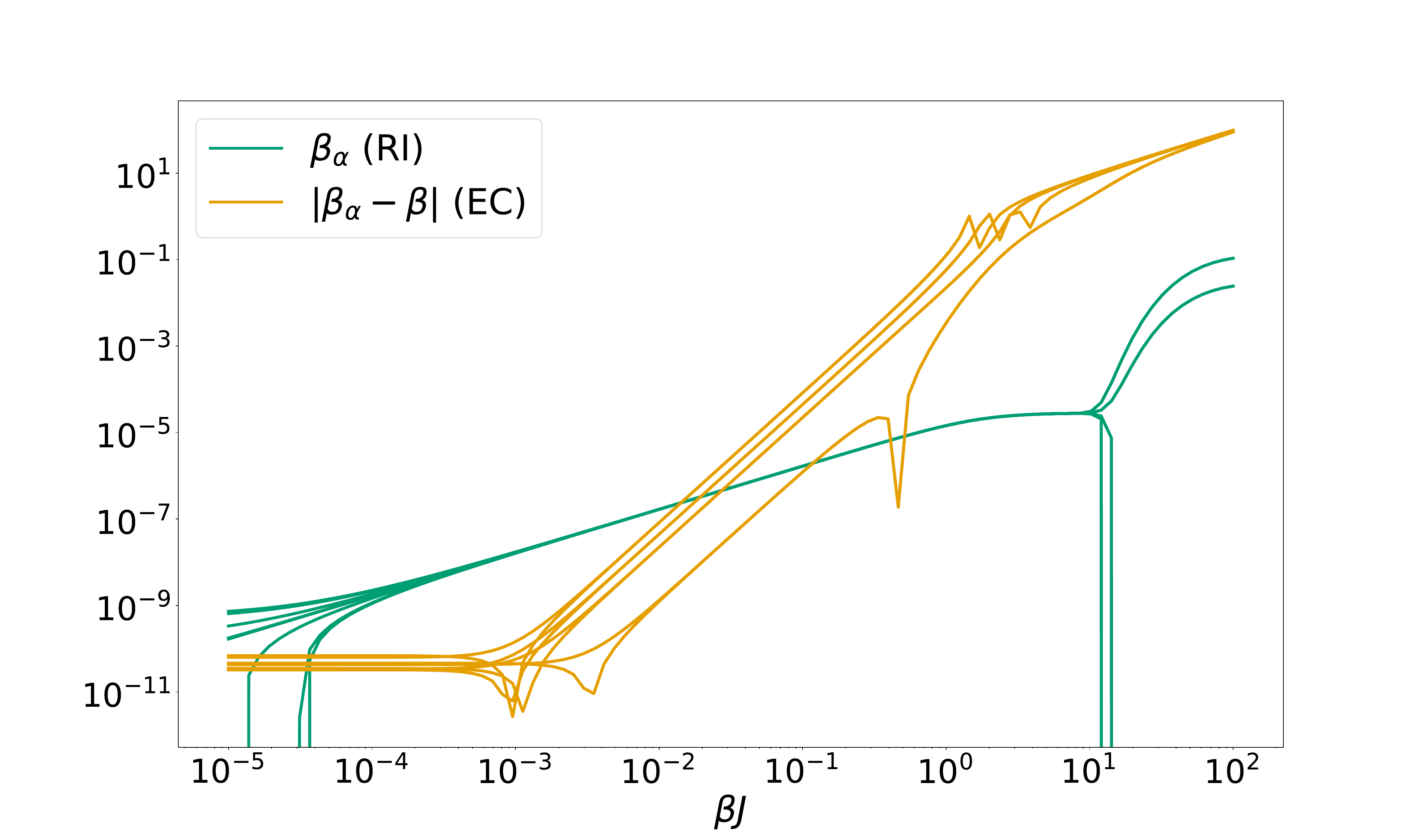}
\caption{\label{fig:betas_scaled} Effective inverse Boltzmann temperatures $\beta_{\alpha}$ corresponding to each diagonal entry of the steady-state single-particle density matrix in the system Hamiltonian's eigenbasis -- as calculated from Eq.~(\ref{eq:fixed_point}) and Eq.~(\ref{eq:effective_fermi_dirac}) for a system of eight sites in a 100-sites-long hopping ring coupled to an environment reset to a thermal state at temperature $\beta$ with period $\tau=0.01$, both in the repeated interactions regime (RI, green) and allowing system-environment correlations to evolve (EC, magenta). Small differences between related lines indicate a Fermi-Dirac distribution of diagonal elements, consistent with a thermal steady state.}
\end{figure}

This is shown in Fig.~\ref{fig:betas_scaled}, revealing an interesting difference in the behaviour of the system depending on whether interactions with the (thermal) environment are modelled by repeated interactions or by repeated resetting where system-environment correlations are allowed to evolve. Whilst the system reaches an approximately thermal state in both cases, only in the latter does it thermalise with the environment \footnote{i.e., evolving towards a thermal state at approximately the same temperature as the environment.}; in the former settling at a state whose temperature is orders of magnitude higher than the environment temperature.

A curious feature of these plots is that they describe approximately thermal steady states above relatively low environment temperatures $\beta J\approx 0.1$, which assuming a coupling between sites of order $J/\hbar\approx 10$ MHz (e.g., as in Ref. \onlinecite{2018PhRvP..10c4050C} for qubit-qubit coupling) corresponds to a temperature of about $T \approx 100$ K. Whilst these temperatures are still too high to be of relevance for many of the current quantum technology components, such as superconducting qubits, it is remarkable that our simple model seems able to describe thermalisation for a fairly wide range of temperatures which include room temperature. There are actually recent proposals of qubits that can operate at elevated temperatures up to ambient temperature
\cite{room_temp_qubit}

\subsection{Spontaneous qubit initialisation\label{subsec:initialisation}}

Initialising a chain of qubits at its ground state is a key example of a control problem of practical importance for quantum technologies, particularly for the implementation of quantum correction codes required for large-scale gate-based quantum computation \cite{2Dcolour_quantcomp,hi_thresh_surf_code,red_decoh}.
One promising approach to this problem is to tune the coupling of this chain to a specially engineered dissipative environment so as to induce fast relaxation to the required state \cite{tunenv,prot_reset,10.1063/5.0166535}. Whilst studies of the speed and fidelity of such protocols have been carried out within a weak-coupling approximation, it has been shown that system-environment interactions play an important role that warrants more accurate analysis \cite{qubit_init}. Therefore, where these environments are to be kept at thermal states throughout this procedure, our method (in the continuous limit) may help to refine studies of their performance.

To completely address this goal is beyond the scope of this paper. However, we can illustrate how this technique is relevant to such studies by applying it to a simplified version of the problem. As such, in what follows we merely ask whether, given a system of linearly coupled qubits, there exists an environment also composed of linearly coupled qubits to which it can be linearly coupled so that, if the environment is maintained at some undefined state, the system evolves towards a pure state where all qubits are in their ground state.

Modelling qubits as lattice sites populated by spinless fermions or hard-core bosons, the Hamiltonian that captures the dynamics in which we are interested can be mapped to the one in Eq. (\ref{eq:quadH}) with $\hat{a}$ a fermionic operator. Moreover, if system indices are chosen to always be lower than environment indices, this implies that states where the system is empty correspond to states where the system is empty. We therefore need only apply the formalism described above to look for attractor solutions corresponding to the system being empty.  

In the language of Eq. (\ref{eq:sODE}), we want to find settings where
$\mathcal{V}_{i}=0\ \forall i$ is an attractor; which is equivalent
to finding settings where $\mathcal{C}_{i}=0\ \forall i$. Without
resetting system-environment correlations, Eq. (\ref{eq:C_cont}) can
be written as
\begin{equation}
\mathcal{C}_{i}=\frac{i}{\hbar}\sum_{\left(\alpha^{\prime},\beta^{\prime}\right)\in\mathcal{E}^{2}}\rho_{\alpha^{\prime}\beta^{\prime}}\left(0\right)\left(M_{\alpha_{i}\alpha^{\prime}}^{\ast}\delta_{\beta_{i}\beta^{\prime}}-\delta_{\alpha_{i}\alpha^{\prime}}M_{\beta_{i}\beta^{\prime}}\right).\label{eq:C_cont_corr}
\end{equation}
Here at least one term inside parenthesis must always be zero (as
$\alpha_{i}$ and $\beta_{i}$ cannot both be environment indices simultaneously),
and therefore we have
\begin{equation}
\mathcal{C}_{i}=\begin{cases}
{i}/{\hbar}\sum_{\alpha^{\prime}\in\mathcal{E}}\rho_{\alpha^{\prime}\beta_{i}}\left(0\right)M_{\alpha_{i}\alpha^{\prime}}^{\ast}, & {\rm if}\ \alpha_{i}\in\mathcal{S}\land\beta_{i}\in\mathcal{E};\\
-{i}/{\hbar}\sum_{\beta^{\prime}\in\mathcal{E}}\rho_{\alpha_{i}\beta^{\prime}}\left(0\right)M_{\beta_{i}\beta^{\prime},} & {\rm if}\ \alpha_{i}\in\mathcal{E}\land\beta_{i}\in\mathcal{S};\\
0, & c.c.
\end{cases}.\label{eq:C_cont_corr_cases}
\end{equation}
 Since $\left(\alpha_{i},\beta_{i}\right)$ runs over all possible
pairs of indices outside of $\mathcal{E}^{2}$, we want to find when
\begin{equation}
\sum_{\beta^{\prime}\in\mathcal{E}}\rho_{\alpha_{E}\beta^{\prime}}\left(0\right)M_{\beta_{S}\beta^{\prime}}=0\ \forall\alpha_{E}\in\mathcal{E}\ \forall\beta_{S}\in\mathcal{S}.\label{eq:reset_cond}
\end{equation}
Choosing to work in the basis where the single-particle density matrix
of the environment is diagonal (i.e., $\rho_{\alpha^{\prime}\beta^{\prime}}\left(0\right)\equiv\delta_{\alpha^{\prime}\beta^{\prime}}n_{\alpha^{\prime}}$),
this boils down to requiring that
\begin{equation}
n_{\alpha_{E}}M_{\beta_{S}\alpha_{E}}=0\ \forall\alpha_{E}\in\mathcal{E}\ \forall\beta_{S}\in\mathcal{S}.
\label{eq:reset_diag}
\end{equation}
This means that the attractor can only be $\mathcal{V}_{i}=0\ \forall i$
if the resetting procedure empties all environment modes which connect
to the system (i.e., $n_{\alpha_{E}}=0\lor M_{\beta_{S}\alpha_{E}}=0\ \forall\alpha_{E}\in\mathcal{E}\ \forall\beta_{S}\in\mathcal{S}$). In other words, the only way this procedure leaves the system exactly empty is if (in this basis) all nodes adjacent to the system are kept empty "by hand" -- or equivalently (as can most easily be seen if those nodes' indices are chosen to be the lowest in the environment), if all corresponding qubits that couple to the system are kept in their ground states. Therefore this simplified approach merely manages to shift the problem to the system's boundary.

\section{Summary and Conclusions\label{sec:Conclusions}}

In this work we have introduced a formalism to study a system in contact with environments that are periodically reset to a fixed arbitrary state, allowing for system-environment correlations to be preserved through the reset. We have also studied the special limit where this period vanishes, which describes an open system with time-independent environments. We have obtained the conditions under which the system eventually loses memory of the initial state, even when the final state is not thermal.

We have applied this formalism to the analysis of a simple system coupled to an environment which is brought to a thermal state at a fixed temperature. In this context we compare the limit where system-environment correlations are erased by the reset (equivalent to the formalism of repeated interactions) to the one where they are allowed to evolve freely; finding that both limits lead to thermal system steady states -- although crucially not for the low temperatures which are relevant for existing quantum technologies. However, only when such correlations evolve freely does the system end up at the same temperature as the environment.

We have further explored the possibilities opened up by this formalism by examining whether a specific environment could be devised that would drive the system to a particular desired state (here corresponding to an initialised chain of qubits). Whilst the answer for this specific example is negative, it illustrates how this approach can naturally contribute to more general problems in environmental engineering.

There are multiple extensions of this work that are facilitated by the mathematical simplicity of quadratic models. Using our reset protocols introduced here, we have recently considered the emergence of coherence-controlled Zeno and anti-Zeno states \cite{kumar2026coherencecontrolledquantumzenodynamics}, interaction-controlled dynamical fixed points \cite{rjlr-7y2k}, and thermodynamics of reset protocols \cite{kumar2026thermodynamicscoherenceselectivequantumreset}. We have further generalized the model to the case of two reservoirs to study memory-transport competition in Fermionic systems \cite{kumar2026memorytransportcompetitionphysicaltwobath}.
Progress has also been made for an exactly solvable model of a qubit interacting with a stream of thermalized ancilla qubits with a Heisenberg-type of Hamiltonian \cite{Prositto_2025}. Similar to the noninteracting case here they find steady states that are diagonal but not necessarily thermalized with the ancilla.

Second, as demonstrated in our second example, this approach may be useful in informing the design of protocols to lead systems to specific states of interest. This way of achieving it is appealing but it is unclear how general steady states can be achieved and what the limitations are. This then appears to be another fruitful direction to follow in future works.

\section*{Data availability}

All data shown in the figures are available from the authors upon reasonable request.

\section*{Acknowledgments}
We wish to thank useful discussions with Jishad Kumar. 
This work has been supported in part by the Academy of Finland through its QTF Center of Excellence grant no. 312298 and by
European Union and the European Innovation Council through the Horizon Europe
project QRC-4-ESP (grant No. 101129663), and EU
Horizon Europe Quest project (No. 10116088).

\bibliography{references}

\newpage

\appendix

\section{Quadratic models} 
\label{sec:quadratic-models}

In general, quadratic models described by Eq.~(\ref{eq:quadH}) can be exactly solved following a change of
variables
\begin{equation}
\hat{a}_{\alpha}=\sum_{n}\psi_{n\alpha}\widetilde{c}_{n},\label{eq:changevar}
\end{equation}
where $\psi_{n\alpha}$ is the $\alpha^{\mathrm{th}}$ element of
the $n^{\mathrm{th}}$ eigenvector of the matrix $M$ (chosen so that
$\sum_{\beta}M_{\alpha\beta}\psi_{n\beta}=\varepsilon_{n}\psi_{n\alpha}$
for some eigenvalue $\varepsilon_{n}$ and $\psi_{n\alpha}$ defines
a unitary matrix), after which the Hamiltonian displays the simple
diagonal form
\begin{equation}
\hat{H}=\sum_{n}\varepsilon_{n}\widetilde{c}_{n}^{\dagger}\widetilde{c}_{n}.\label{eq:diagH}
\end{equation}
 Owing to the unitarity of $\psi_{n\alpha}$, Eq. (\ref{eq:changevar})
can be inverted to yield
\begin{equation}
\widetilde{c}_{n}=\sum_{\alpha}\psi_{n\alpha}^{\ast}\hat{a}_{\alpha},\label{eq:inv_changevar}
\end{equation}
where $\widetilde{c}_{n}^{\dagger}$ and $\widetilde{c}_{n}$ are,
respectively, creation and annihilation operators of the same kind
as their counterparts in Eq. (\ref{eq:quadH}) -- and therefore evolution according to the Hamiltonian in Eq. (\ref{eq:diagH}) can be linearly mapped to the dynamics of $N$ free particles
of energies given by $\varepsilon_{n}$.

The state of a system described by the Hamiltonian in Eq. (\ref{eq:quadH})
is fully determined by the single-particle density matrix $\rho_{\alpha\beta}$,
which can be written as
\begin{equation}
\rho_{\alpha\beta}\equiv\left\langle \hat{a}_{\alpha}^{\dagger}\hat{a}_{\beta}\right\rangle =\sum_{nm}\psi_{n\alpha}^{\ast}\psi_{m\beta}\widetilde{\rho}_{nm},\label{eq:spdm}
\end{equation}
where $\widetilde{\rho}_{nm}$ is the single-particle density matrix
defined with respect to the diagonalised Hamiltonian in Eq. (\ref{eq:diagH}),
\begin{equation}
\widetilde{\rho}_{nm}\equiv\left\langle \widetilde{c}_{n}^{\dagger}\widetilde{c}_{m}\right\rangle =\sum_{\alpha\beta}\psi_{n\alpha}\psi_{m\beta}^{\ast}\rho_{\alpha\beta}.\label{eq:til-spdm}
\end{equation}

The unitary dynamics of $\widetilde{\rho}_{nm}$ is just given by
\begin{equation}
\widetilde{\rho}_{nm}\left(t\right)=\widetilde{\rho}_{nm}\left(0\right)e^{i\left(\varepsilon_{n}-\varepsilon_{m}\right)t/\hbar},\label{eq:time_til_spdm}
\end{equation}
 where $t$ represents time and $\hbar$ is the reduced Planck constant.
Changing back to the original variables this translates to the relation
\begin{equation}
\rho_{\alpha\beta}\left(t\right)=\sum_{\alpha^{\prime}\beta^{\prime}}\rho_{\alpha^{\prime}\beta^{\prime}}\left(0\right)\sum_{nm}\psi_{n\alpha}^{\ast}\psi_{m\beta}\psi_{n\alpha^{\prime}}\psi_{m\beta^{\prime}}^{\ast}e^{{i}\left(\varepsilon_{n}-\varepsilon_{m}\right)t/\hbar},\label{eq:spdm_evol}
\end{equation}
which can be equivalently written as
\begin{equation}
\rho_{\alpha\beta}\left(t\right)=\sum_{\alpha^{\prime}\beta^{\prime}}U_{\alpha\alpha^{\prime}}^{\ast}\left(t\right)U_{\beta\beta^{\prime}}\left(t\right)\rho_{\alpha^{\prime}\beta^{\prime}}\left(0\right),\label{eq:spdm_evol_op}
\end{equation}
where we have introduced the components of the evolution operator
\begin{equation}
\begin{split}
U_{\alpha\beta}\left(t\right)
    &\equiv\left\langle \alpha\right|\hat{U}\left(t\right)\left|\beta\right\rangle \equiv\left\langle \alpha\right|e^{-i\hat{H}t/\hbar}\left|\beta\right\rangle\\
    &=\left(e^{-iMt/\hbar}\right)_{\alpha\beta}=\sum_{n}e^{-i\varepsilon_{n}t/\hbar}\psi_{n\alpha}\psi_{n\beta}^{\ast},
\end{split}
\label{eq:evol_op}
\end{equation}
$\left|\alpha\right\rangle \equiv\hat{a}_{\alpha}^{\dagger}\left|0\right\rangle $
being an element of this system's Fock state, and $\left|0\right\rangle $
its vacuum.


\section{Fixed point of Eq.~(\ref{eq:lineq})} 
\label{sec:uniqueness_of_fixed_point_of_eq_eq:lineq}

Equation (\ref{eq:lineq}) can always be solved if the matrix $P$ that diagonalises
$D$ (i.e., such that $\sum_{k}\sum_{a}P_{ik}D_{ka}P_{ac}^{-1}=\lambda_{i}\delta_{ic}$,
where $\lambda_{i}$ are the eigenvalues of $D$) is known \footnote{$D$ must be diagonalisable at least for sufficiently small $\tau$
due to it being Hermitian for $\tau=0$ and continuous in $\tau$.}. Changing to the eigenbasis of $D$,
\begin{equation}
\widetilde{V}_{i}\left[n\right]\equiv\sum_{j}P_{ij}V_{j}\left[n\right],\label{eq:eigenvector}
\end{equation}
and straightforwardly manipulating Eq. (\ref{eq:lineq})
\[
\widetilde{V}_{i}\left[n+1\right]=\sum_{k}P_{ik}\left(\sum_{j}D_{kj}V_{j}\left[n\right]+C_{k}\right)
\]
\[=\sum_{k}\sum_{j}P_{ik}D_{kj}V_{j}\left[n\right]+\sum_{k}P_{ik}C_{k}
\]
\[
=\sum_{k}\sum_{a}\sum_{b}P_{ik}D_{ka}\delta_{ab}V_{b}\left[n\right]+\sum_{k}P_{ik}C_{k}
\]
\[=\sum_{k}\sum_{a}\sum_{b}P_{ik}D_{ka}\sum_{c}P_{ac}^{-1}P_{cb}V_{b}\left[n\right]+\sum_{k}P_{ik}C_{k}
\]
\[
=\sum_{k}\sum_{a}\sum_{b}\sum_{c}P_{ik}D_{ka}P_{ac}^{-1}P_{cb}V_{b}\left[n\right]+\sum_{k}P_{ik}C_{k}
\]
\begin{equation}
=\sum_{b}\sum_{c}\lambda_{i}\delta_{ic}P_{cb}V_{b}\left[n\right]+\sum_{k}P_{ik}C_{k},\label{eq:manipulate}
\end{equation}
we find the simple relation
\begin{equation}
\widetilde{V}_{i}\left[n+1\right]=\lambda_{i}\widetilde{V}_{i}\left[n\right]+\widetilde{C}_{i},\label{eq:eigenlineq}
\end{equation}
 where
\begin{equation}
\widetilde{C}_{i}=\sum_{k}P_{ik}C_{k}.\label{eq:Ctil}
\end{equation}

Solutions to Eq. (\ref{eq:eigenlineq}) are of the form
\begin{equation}
\widetilde{V}_{i}\left[n\right]=\begin{cases}
\left(\widetilde{V}_{i}\left[0\right]+\frac{\widetilde{C}_{i}}{\lambda_{i}-1}\right)\lambda_{i}^{n}-\frac{\widetilde{C}_{i}}{\lambda_{i}-1}, & {\rm if\ }\lambda_{i}\neq1;\\
\widetilde{V}_{i}\left[0\right]+n\widetilde{C}_{i}, & {\rm if\ }\lambda_{i}=1,
\end{cases}\label{eq:fin_sol}
\end{equation}
from which we can conclude that physical solutions must have $\left|\lambda_{i}\right|\leq1$
and $\lambda_{i}=1\Rightarrow\widetilde{C}_{i}=0$ for all $i$ (lest the
correlation functions diverge in the infinite future) and that if
$\lambda_{i}\neq1$ then the only fixed point is the state vector 
\begin{equation}
\widetilde{V}_{i}=-\frac{\widetilde{C}_{i}}{\lambda_{i}-1},\label{eq:fixed_point}
\end{equation}
which is actually an attractor if $\left|\lambda_{i}\right|<1$ (and
an attractor for the time average after long times for all physical
solutions). Therefore, we can also see that the issue of pseudo-thermalisation depends solely on properties of the Hamiltonian -- as these models necessarily pseudo-thermalise if $\lambda_{i}\neq 1 \forall i$.


\section{Fixed point in the continuous limit} 
\label{sec:uniqueness_of_fixed_point_in_the_continuous_limit}

For small $\tau$, Eq.~(\ref{eq:U_tauexpand}) can be substituted into Eqs. (\ref{eq:D}) and (\ref{eq:C}) to find
\begin{equation}
D_{ij}=\delta_{ij}+\frac{i\tau}{\hbar}\left(M_{\alpha_{i}\alpha_{j}}^{\ast}\delta_{\beta_{i}\beta_{j}}-\delta_{\alpha_{i}\alpha_{j}}M_{\beta_{i}\beta_{j}}\right)+\mathcal{O}\left(\tau^{2}\right)\label{eq:D_tauexpand},
\end{equation}
and
\begin{equation}
\begin{split}
C_{i}
    &=\frac{i\tau}{\hbar}\sum_{\left(\alpha^{\prime},\beta^{\prime}\right)\in\mathcal{R}}\rho_{\alpha^{\prime}\beta^{\prime}}\left(0\right)\left(M_{\alpha_{i}\alpha^{\prime}}^{\ast}\delta_{\beta_{i}\beta^{\prime}}-\delta_{\alpha_{i}\alpha^{\prime}}M_{\beta_{i}\beta^{\prime}}\right)\\
    &+\mathcal{O}\left(\tau^{2}\right).
\end{split}
\label{eq:C_expandtau}
\end{equation}

In the extreme version of this limit, when $\tau=0$ and thus each
iterative transformation is only allowed to differ from identity infinitesimally,
a continuous approach is required. This can be straightforwardly done
by defining the continuous variable
\begin{equation}
\mathcal{V}_{i}\left(t\right)\equiv\rho_{\alpha_{i}\beta_{i}}\left(t\right)=\lim_{\tau\rightarrow0}V_{i}\left[\frac{t}{\tau}\right]\label{eq:V_cont}
\end{equation}
and applying Eq. (\ref{eq:lineq}) to write the linear system of ordinary
differential equations
\begin{equation}
\frac{d\mathcal{V}_{i}}{dt}\left(t\right)=\sum_{j}\mathcal{D}_{ij}\mathcal{V}_{j}\left(t\right)+\mathcal{C}_{i},\label{eq:sODE}
\end{equation}
where we have introduced the new matrix
\begin{equation}
\mathcal{D}_{ij}\equiv\lim_{\tau\rightarrow0}\frac{D_{ij}-\delta_{ij}}{\tau}=\frac{i}{\hbar}\left(M_{\alpha_{i}\alpha_{j}}^{\ast}\delta_{\beta_{i}\beta_{j}}-\delta_{\alpha_{i}\alpha_{j}}M_{\beta_{i}\beta_{j}}\right)\label{eq:D_cont},
\end{equation}
and the new vector
\begin{equation}
\begin{split}
\mathcal{C}_{i}
    &\equiv\lim_{\tau\rightarrow0}\frac{C_{i}}{\tau}\\
    &=\frac{i}{\hbar}\sum_{\left(\alpha^{\prime},\beta^{\prime}\right)\in\mathcal{R}}\rho_{\alpha^{\prime}\beta^{\prime}}\left(0\right)\left(M_{\alpha_{i}\alpha^{\prime}}^{\ast}\delta_{\beta_{i}\beta^{\prime}}-\delta_{\alpha_{i}\alpha^{\prime}}M_{\beta_{i}\beta^{\prime}}\right).
\end{split}
\label{eq:C_cont}
\end{equation}

Since it follows from Eq. (\ref{eq:D_cont}) that any matrix $P$ that
diagonalises $D$ must also diagonalise $\mathcal{D}$, we can solve
Eq. (\ref{eq:sODE}) analogously to Eq. (\ref{eq:lineq}), finding
\begin{equation}
\widetilde{\mathcal{V}}_{i}\left(t\right)=\begin{cases}
e^{\sigma_{i}t}\left(\widetilde{\mathcal{V}}_{i}\left(0\right)+{\widetilde{\mathcal{C}}_{i}}/{\sigma_{i}}\right)-{\widetilde{\mathcal{C}}_{i}}/{\sigma_{i}}, & \mathrm{if}\ \sigma_{i}\neq0;\\
\widetilde{\mathcal{V}}_{i}\left(0\right)+\widetilde{\mathcal{C}}_{i}t, & \mathrm{if}\ \sigma_{i}=0,
\end{cases}\label{eq:fin_sol_sODE}
\end{equation}
where, as before, tilde denotes multiplication by $P$, and $\sigma_{i}=\left.d\lambda_{i}/d\tau\right|_{\tau=0}=\lim_{\tau\rightarrow0}\tau^{-1}\ln\lambda_{i}$
are the eigenvalues of $\mathcal{D}$ (so that$\sum_{k}\sum_{a}P_{ik}\mathcal{D}_{ka}P_{ac}^{-1}=\sigma_{i}\delta_{ic}$).
Also analogously to the discrete case, we can conclude that physical
solutions must have $\rm{Re}\left(\sigma_{i}\right)\leq0$ and $\sigma_{i}=0\Rightarrow\widetilde{\mathcal{C}}_{i}=0$
for all $i$ and that if $\sigma_{i}\neq0$ then the only fixed point
is the initial vector 
\begin{equation}
\mathcal{V}_{i}\left(0\right)=-\frac{\widetilde{\mathcal{C}}_{i}}{\sigma_{i}},\label{eq:fixed_point_cont}
\end{equation}
which is an attractor as long as $\operatorname{Re}(\sigma_{i})<0$.
We note that this is analogous to the quantum Zeno effect, where continuous measurement of the system halts temporal evolution of the system \cite{Facchi_2008}.
\bigskip

\section{Infinite temperature attractor} 
\label{sec:inf_temp}

Consider an arbitrary quadratic
model described by a Hamiltonian of the form in Eq. (\ref{eq:quadH}), where the indices in some set $\mathcal{S}$ are interpreted
as corresponding to a system of interest and the indices in its complement
$\mathcal{E}$ are interpreted as corresponding to an environment.
At each iteration the state of the environment is reset to a thermal
state at infinite temperature, i.e.,
\begin{equation}
\rho_{\alpha\beta}\left(N\tau\right)=\frac{1}{2}\delta_{\alpha\beta}\, \, \forall\alpha,\beta\in\mathcal{E}\forall N\in\mathbb{Z}.\label{eq:inf_temp}
\end{equation}

Regardless of whether system-environment correlations are reset to
zero or allowed to evolve, Eq. (\ref{eq:C}) can then be written as
\begin{equation}
C_{i}=\frac{1}{2}\sum_{\alpha^{\prime}\in\mathcal{E}}U_{\alpha_{i}\alpha^{\prime}}^{\ast}\left(\tau\right)U_{\beta_{i}\alpha^{\prime}}\left(\tau\right),\label{eq:C_inf_temp}
\end{equation}
which allows us to see that one of the fixed points of the system
must always correspond to the infinite temperature state (with vanishing
system-environment correlations), i.e.,
\begin{equation}
V_{i}=\frac{1}{2}\delta_{\alpha_{i}\beta_{i}} \, \, \forall i,\label{eq:Vi_inf_temp}
\end{equation}
as then substitution into Eq. (\ref{eq:D}) yields
\begin{equation}
\sum_{j}D_{ij}V_{j}=\frac{1}{2}\sum_{\alpha^{\prime}\in\mathcal{S}}U_{\alpha_{i}\alpha^{\prime}}^{\ast}\left(\tau\right)U_{\beta_{i}\alpha^{\prime}}\left(\tau\right)\label{eq:D_mult_inf_temp}
\end{equation}
and thus
\begin{equation}
\sum_{j}D_{ij}V_{j}+C_{i}=\frac{1}{2}\sum_{\alpha^{\prime}}U_{\alpha_{i}\alpha^{\prime}}^{\ast}\left(\tau\right)U_{\beta_{i}\alpha^{\prime}}\left(\tau\right)=\frac{1}{2}\delta_{\alpha_{i}\beta_{i}}=V_{i}.\label{eq:inf_temp_identity}
\end{equation}

Therefore generic quadratic models are expected to thermalise with any environment kept at infinite temperature.


\end{document}